\documentstyle[prl,aps]{revtex}
\begin{document}

%\preprint{Preprint} \draft{}

% rein...
\twocolumn[\hsize\textwidth\columnwidth\hsize
           \csname @twocolumnfalse\endcsname

\title{Observation of the critical regime near Anderson localization of light}

\author{Martin St\"orzer, Peter Gross, Christof M. Aegerter, and Georg Maret}
\address{Fachbereich Physik, University of Konstanz, Germany}
\date{\today}
\maketitle \widetext

\begin{abstract}
Diffusive transport is among the most common phenomena in nature
\cite{eini}. However, as predicted by Anderson \cite{Anderson},
diffusion may break down due to interference. This transition from
diffusive transport to localization of waves should occur for any
type of classical or quantum wave in any media as long as the
wavelength becomes comparable to the transport mean free path
$\ell^*$ \cite{Ioffe}. The signatures of localization and those of
absorption, or bound states, can however be similar, such that an
unequivocal proof of the existence of wave localization in
disordered bulk materials is still lacking. Here we present
measurements of time resolved non-classical diffusion of visible
light in strongly scattering samples, which cannot be explained by
absorption, sample geometry or reduction in transport velocity.
Deviations from classical diffusion increase strongly with
decreasing $\ell^*$ as expected for a phase transition. This
constitutes an experimental realization of the critical regime in
the approach to Anderson localization.
\end{abstract}

\pacs{DRAFT VERSION: NOT FOR DISTRIBUTION}

\narrowtext ]

Diffusive transport controls many phenomena in biology, chemistry,
physics and engineering, where undirected transport is
characterized by a linear increase of the mean square displacement
$\langle r^2 \rangle$ with time \cite{eini}. As a consequence,
transmission of particles through a slab of length $L$ is
proportional to $\ell^*/L$ known in the context of electrons in
metals as Ohm's law. However, when the wave-nature of the
diffusing particles is taken into account, constructive
interference of waves propagating on reciprocal multiple
scattering paths may lead to a breakdown of diffusion and the
particles are "trapped" on closed multiple scattering loops. This
means that the probability of returning to the starting point of
such a closed loop is increased twofold due to the fact that the
path has the same length in both counter-propagating directions.
Thus diffusive transport away from this starting point is reduced.
Anderson first predicted this destruction of diffusion in 1958
\cite{Anderson} to explain the metal-insulator transition. If the
scattering power of the medium is high, i.e. the transport mean
free path $\ell^*$ is small, the density of closed loops increases
which leads to a reduced transport inside the material. This can
be described by a rescaling of the diffusion coefficient
\cite{Abrahams}. For strong localization the rescaled diffusion
coefficient becomes zero at finite length scales, such that the
spread of $\langle r^2 \rangle$ comes to an end at a length scale
corresponding to the localization length. The transition from a
diffusive to a localized state should occur when the wavelength
$\lambda$ becomes comparable to $\ell^*$, as quantified by the
Ioffe-Regel criterion $k \ell^* \simeq 1$ where $k = 2\pi/\lambda$
is the wavenumber \cite{Ioffe}.

While localization of electrons in metals was widely studied
\cite{altshuler}, an unequivocal demonstration of the pure
interference effect as predicted by Anderson turned out to be
difficult in this case, as electrons bound in deep minima of a
random potential are virtually impossible to distinguish from
those in closed loops. However, photons in disordered materials
revealed an excellent model system to study localization effects
\cite{Anderson2,john}: In contrast to electrons, they do not
interact with each other, nor can they be bound in a random
potential. Hence all deviations from classical diffusion have to
be due to interference, when absorption is taken into account
properly. Moreover, in order to create a strongly scattering
medium, materials having a large optical refractive index combined
with very small absorption are available for visible light.

Thus there have been many investigations in the past studying
diffusive transport of visible \cite{schuurmans1,schuurmans2} and
infrared light \cite{rivas}. For visible light, where reported
values of $k \ell^*$ go down to 3.2 \cite{schuurmans1},
measurements have mainly focused on static transmission
experiments where a decay faster than $\propto 1/L$ was
interpreted as evidence for the onset of localization
\cite{wiersma}. It was noted however that absorption, which is
always present in such samples, can lead to a similar decay
\cite{schefold,AG1}. Furthermore, recent investigations of time
resolved transmission of these samples with similar values of $k
\ell^*$ show no discernible deviation from purely diffusive
behavior \cite{schuurmans2}. Localization of photons has also been
studied by use of microwave scattering in disordered media
\cite{AG1,AG2,AG3}, where indications of a decreasing diffusion
coefficient have been found \cite{AG3}. However it has to be noted
that these investigations were carried out in tubes of extent
$\sim 2 \times 2 \times 20$ in units of $\ell^*$ \cite{cheung},
which constitute a relatively small quasi one-dimensional (1D)
system. As localization is always present in 1D and 2D
\cite{Abrahams}, theoretical descriptions of the quasi 1D geometry
\cite{cheung,skipetrov1} indicate that observations of
localization effects are possible far away from $k\ell^* \simeq
1$. Similar theories in 3D \cite{skipetrov2} find a negligible
effect in the same regime. Thus, an unequivocal proof of the
transition to localization in three dimensions is still lacking.

Here we present measurements of time resolved photon transport
through bulk powders of TiO$_2$ with typical sample dimensions of
$\sim 10^5 \times 10^5 \times 10^4$ in units of $\ell^*$. These
measurements have the advantage that they allow a direct
determination of the photon path length dependence of the
diffusion coefficient, where absorption and localization lead to
different functional forms of the tail (exponential vs.
non-exponential) of the path length distribution
\cite{skipetrov1}. Furthermore we independently measure the
turbidity of the samples using coherent backscattering
\cite{Albada,Wolf}. This allows a systematic study of the approach
to the localization transition as a function of turbidity
$(k\ell^*)^{-1}$.

Our samples consist of ground TiO$_2$ particles in its rutile
structure with a refractive index of 2.8. The resulting grains of
the different samples have various average particle sizes with
average diameters ranging from 220 nm to 550 nm with a
polydispersity of $\sim 25 \%$. Those particles are commercially
available as pigments for white paint from e.g. DuPont or Aldrich.
In order to minimize $\ell^*$ these powders are compressed to
obtain filling fractions of $\phi \simeq 0.4$. The scattering
properties are determined from coherent backscattering
\cite{Albada,Wolf}. The angular width of this enhancement to the
incoherent background is inversely proportional to $k \ell^*$
\cite{Akkerman,Tigelen}. For the highly scattering samples studied
here, the width becomes very broad, which is why the distribution
has to be measured to very wide angles for a good determination of
the incoherent background. Our setup was custom designed for this
problem (see methods) \cite{peter}.

\begin{figure}[hbt]
\input{epsf}
\epsfxsize 10cm
\centerline{\epsfbox{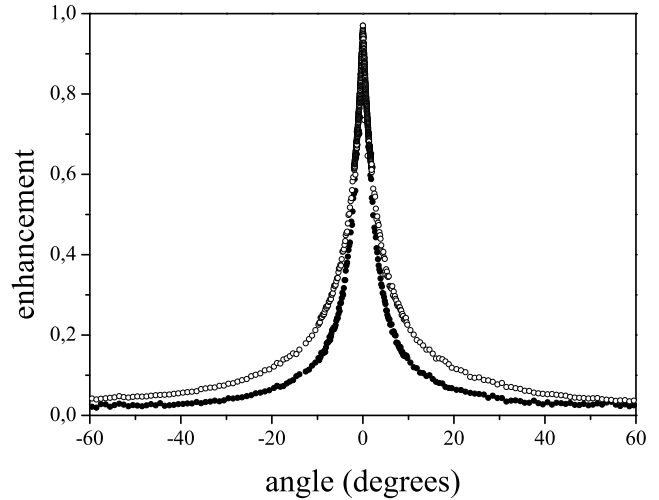}} \caption[~]{Measurements
of coherent backscattering for two different samples. Open
symbols: sample S1 with an average particle diameter of 250 nm
which yields $k \ell^* = 2.5$. Closed symbols: sample S3 with an
average diameter of 550 nm and $k \ell^* = 6.3$. All measurements
were done with circularly polarized light at a wavelength $2\pi /k
= 590$~nm.} \label{Cone}
\end{figure}

Fig.\ref{Cone} shows the backscattering angular distribution for
two different samples, S1 and S3, corresponding to average
particle diameters of 250 nm and 550 nm respectively. Using the
value of the averaged refractive index from numerical simulations
based on the energy density coherent potential approximation
\cite{Garnett}, we determine the reflectivity of the surface and
are hence able to correct for the overestimation of $k\ell^*$ due
to internal reflections \cite{Lstern}. Thus the values of $k
\ell^*$ are determined from the full widths at half maximum of the
curves with a correction for the reflectivity of the surface. This
gives values ranging from $k \ell^*=6.3$ for S3 to $k \ell^*=2.5$
for S1 at a wavelength of 590~nm, close to the Ioffe-Regel
criterion.

\begin{figure}[hbt]
\input{epsf}
\epsfxsize 8cm
\centerline{\epsfbox{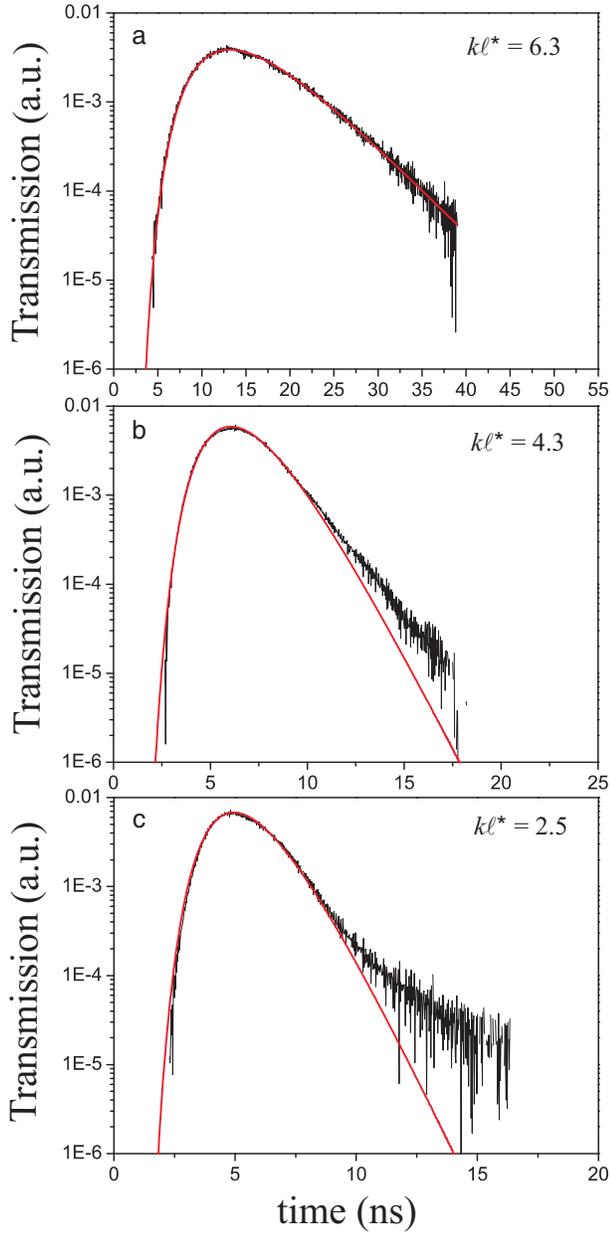}} \caption[~]{Path length
distributions from samples S1, S2, and S3. The experimental
results are compared to diffusion theory including absorption. In
A, one can see that the data from S3 ($L = 2.5$ mm, $D = 22$
m$^2$/s, $\ell_a = 2600$ mm) closely follow the diffusion fit,
showing an exponential decay at long times. Part C in contrast
shows strong deviations from the diffusion fit for S1 ($L = 1.48$
mm, $D = 15$ m$^2$/s, $\ell_a = 340$ mm), with a clearly
non-exponential decay at long times. These deviations can be
explained by a time dependent diffusion coefficient in the sample.
An intermediate case is shown in part B from sample S2 ($L = 1.51$
mm, $D = 13$ m$^2$/s, $\ell_a = 380$ mm), with a value of $k\ell^*
= 4.3$, where small deviations from the classical behaviour can be
observed.} \label{TOF}
\end{figure}

In order to measure the time resolved transmission we use a single
photon counting method, where the time delay of a picosecond light
pulse transmitted through the sample is measured. From a histogram
of time of flights of many such pulses, the path length
distribution inside the sample is obtained directly
\cite{schuurmans2,Watson,drake}. Our setup consists of a dye laser
working at a wavelength of 590 nm with a pulse width of $\sim$20
ps. The time of flight histograms have been deconvoluted with the
pulse shape of the laser system in order to recover the pure path
length distributions (see methods). Fig.~\ref{TOF} shows the path
length distributions in the samples characterized in
Fig.~\ref{Cone} compared to another sample (S2) with $k \ell^*
=4.3$. As can be seen in Fig.~\ref{TOF}A, classical diffusion
theory including absorption \cite{Lenke,Berkovitz} fits the data
from sample S3 very well. In contrast, the path length
distribution of S1 (Fig.~\ref{TOF}C) shows marked deviations from
diffusive behaviour. In particular, we observe a non-exponential
decay at long times, with photons staying inside the sample longer
than expected from a purely diffusive process. In Fig.~\ref{TOF}B,
the data from sample S2 show small deviations from the diffusion
picture.

\begin{figure}[hbt]
\input{epsf}
\epsfxsize 10cm
\centerline{\epsfbox{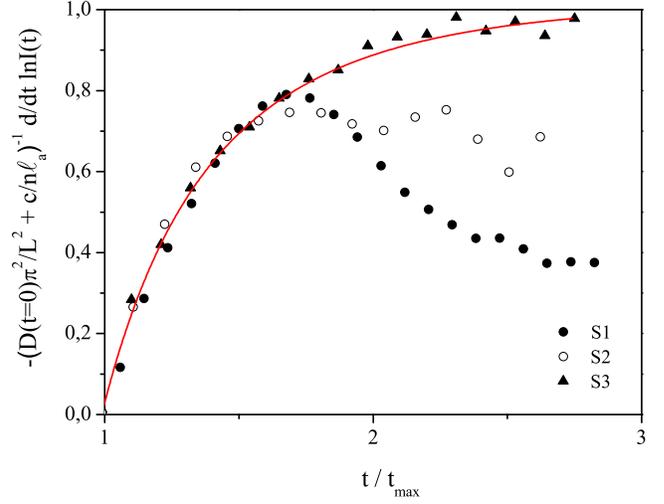}} \caption[~]{Taking the
negative derivative of the logarithm of the measured data one
obtains a measure of the time dependence of the diffusion
coefficient out of the time of flight data. Sample S3 has a
constant diffusion coefficient at long times in agreement with
diffusion theory indicated by the solid line. S1 and S2 in
contrast show a decay illustrating the decrease of the diffusion
coefficient of light with time. This demonstrates the existence of
localized modes. } \label{Diff}
\end{figure}

Diffusion in an extended slab of thickness $L$, which in our
experiments ranges from 1.3 mm to 2.5 mm, leads to an exponential
decay of the time dependent transmission at long times as $I(t)
\propto e^{- \left(\frac{\pi^2 D(t)}{L^2}+\frac{c}{n
\ell_a}\right)t}$ \cite{Lenke,Berkovitz}. Here $\ell_a$ is the
absorption length, which in our samples ranges from 0.3 m to 2.6
m, $c$ is the speed of light in vacuum, $n$ is the effective
refractive index of the medium, and $D$ the diffusion coefficient.
Note that the values for $\ell_a$ are more than a factor of $10^6$
larger than typical values of $\ell^*$ in our samples. Any
non-exponential decay of $I(t)$ at long times thus indicates a
temporally varying diffusion coefficient $D(t)$. Using the
properties of $I(t)$ at long times, we can obtain a direct
measurement of the diffusion coefficient up to a constant given by
$\ell_a$. This is achieved by taking the negative time derivative
of the logarithm of the intensity divided by ($\pi^2D(t=0)/L^2 +
c/n \ell_a$), where $\pi^2D(t=0)/L^2$ is the inverse of the time
$t_{max}$, at which most photons leave the sample \cite{AG3}. This
is shown in Fig.~\ref{Diff} for samples S1, S2, and S3. For long
times the classical sample S3 approaches the constant value of 1,
whereas the curve for S1 shows a strongly decreasing diffusion
coefficient and S2 reveals intermediate behaviour.

\begin{figure}[hbt]
\input{epsf}
\epsfxsize 10cm
\centerline{\epsfbox{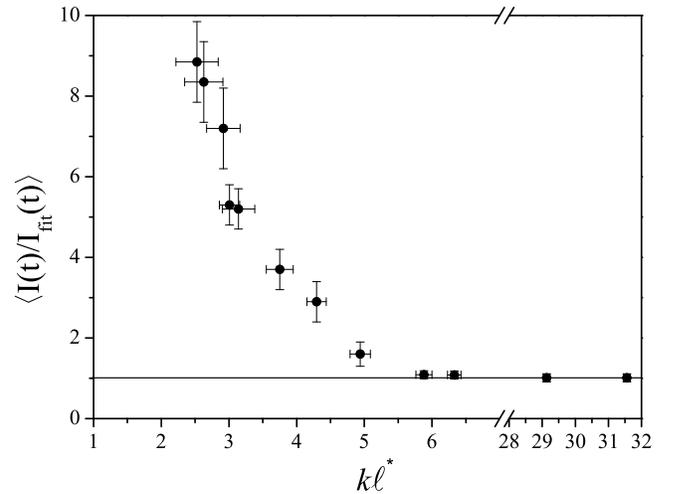}} \caption[~]{Measurements
of the deviations from diffusion on several samples with different
values of $k\ell^*$. The deviation is quantified by taking the
ratio of the experimental path length distribution, $I(t)$, to the
classical fit, $I_{fit}(t)$, and averaging it over the time
interval from $t_{max}$ to $3 t_{max}$. The results were plotted
over the $k \ell^*$ value determined from coherent backscattering
(see text). Samples that have a $k \ell^*$ nearer to the
Ioffe-Regel criterion show a bigger deviation from diffusive
behaviour. } \label{Local}
\end{figure}

To quantify the non-classicality of the diffusive light transport
through the different samples we determined the average of the
ratio of the measured data to the diffusion fits. This was done
systematically for all powders over the time interval from
$t_{max}$ to $3 t_{max}$. This deviation is given as a function of
$k \ell^*$ in Fig. \ref{Local}. The figure clearly shows that the
deviations increase strongly when approaching the Ioffe-Regel
criterion, as expected for a phase transition.

The data we have presented show clear deviations from diffusive
transport through these highly scattering samples. The
non-exponential decay of the path length distribution indicates a
renormalized value of the diffusion coefficient at long times as
was predicted by scaling theory \cite{Abrahams,john}. Moreover,
the deviations from diffusive behaviour scale with the value of $k
\ell^*$ consistent with the Ioffe-Regel criterion \cite{Ioffe} and
the approach to a phase transition. These deviations cannot be
explained by absorption as this only leads to an additional
exponential decrease but cannot introduce a non-exponential path
length distribution. There is also no systematic dependence of the
values of the absorption length of the various samples on $k
\ell^*$ in contrast to the monotonic dependence of the deviations
from classical diffusion. A change in the transport velocity due
to resonant scattering \cite{albada2} can also be ruled out, since
values of the diffusion coefficient from the classical fits, such
as that in Fig. \ref{TOF}C, are consistent with a value of the
transport velocity of $c/n$. Furthermore such a change is not
expected to be dependent on the length of the specific path and
hence would not lead to a non-exponential decay as the data show.
Finally, sample inhomogeneities, such as a stratification, cannot
account for the effect, as the path length distributions are found
to be independent of the direction of illumination. We thus
conclude that these observations constitute direct evidence for
the slowing down of photon diffusion due to the approach to the
Anderson localization transition and hence for the existence of
the transition to strong localization of photons in three
dimensions.
\\
\section*{Methods}
\subsection*{Coherent backscattering}
The diffuse reflected intensity is not constant but depends on the
backscattering angle $\theta$ as $\cos(\theta)$, such that the
difference of the backscattering signal from this functional shape
has to be determined. Our setup consists of 256 photo-sensitive
diodes attached to an arc with a diameter of 1.2~m in order to get
sufficient angular resolution over a range of $-70^\circ < \theta
< +70^\circ$ \cite{peter}. Here, the resolution is 1$^\circ$ for
$|\theta|>20^\circ$, 0.7$^\circ$ for $10^\circ<|\theta|<20^\circ$
and 0.14$^\circ$ for $|\theta|<10^\circ$. In addition, the central
part of the backscattering cone, $-3^\circ < \theta < +3^\circ$
was measured separately using a beam-splitter and a charged
coupled device camera to a resolution of 0.02$^\circ$. The
measurements are done using circularly polarized light in order to
reduce the influence of singly scattered light.

\subsection*{Time of flight measurements}
Our setup consists of a dye laser working at a wavelength of 590
nm with a pulse width of $\sim$20 ps. The Rhodamin6G dye laser
(Coherent C699) is pumped by an Ar$^+$ laser (Coherent Inova400)
and was modified by APE with a mode locker and a cavity dumper to
deliver picosecond pulses. In order to recover the pure path
length distributions, the time of flight histograms have to be
deconvoluted with the pulse shape of the laser system. This is
because after pulses, albeit strongly suppressed, and
indiscriminate noise may lead to disturbances in the time of
flight measurements. For this purpose, we measured pulse shape,
including the noise level, in the absence of a sample. This
zero-pulse was then deconvoluted with the time of flight data in
Fourier space to directly give the path length distribution for a
supposed delta-peaked pulse as it is calculated theoretically.

The observed non-exponential decay is not due to a distribution of
$\ell^*$ values in the sample due to e.g. stratification. We have
confirmed this by measuring the time of flight distributions with
sample S1 flipped, such that the direction of stratification would
be reversed. These measurements show the same non-exponential
decay as Fig.~\ref{TOF}C.

\section*{Acknowledgements}

This work was supported by the Deutsche Forschungsgemeinschaft,
the International Research and Training Group "Soft Condensed
Matter of Model Systems" and the Center for Applied Photonics
(CAP) at the University of Konstanz. Furthermore, we would like to
thank DuPont chemicals and Aldrich for providing samples used in
this study.
\bibliographystyle{prsty}

%\end{multicols}
\end{document}